\documentclass[runningheads]{llncs}
\usepackage{graphicx}
\usepackage{multirow}
\usepackage{rotating}
\usepackage{tabularx}
\usepackage{amsmath}
\usepackage{todonotes}
\usepackage{xcolor}
\usepackage{tablefootnote} 
\definecolor{bluegreen}{RGB}{0,153,153}
\definecolor{bluepurple}{RGB}{102,0,204}
\definecolor{bluered}{RGB}{204,0,102}

\usepackage{algorithm}  
\usepackage{algorithmicx}
\usepackage{algpseudocode}
\usepackage{subfigure}
\usepackage{adjustbox}
\usepackage{blindtext}
\usepackage{tabularx}

\newcommand{\SSName}{RescureService} 
\newcommand{\ScenarioName}{disaster relief scenario}  

\begin{document}
\title{\SSName: A Benchmark Microservice System for the Research of Mobile Edge and Cloud Computing}


%
\titlerunning{\SSName: A Benchmark Microservice System}
%
\author{Xiang He \and Teng Wang \and Lei Liu \and Jianan Li \and Zihang Su \and Yingming Guo \and Zhiying Tu \and Hanchuan Xu \and Zhongjie Wang}
%
%
\institute{Faculty of Computing, Harbin Institute of Technology, Harbin, China
\email{hexiang@hit.edu.cn, \{21S003068, liulei, 20S003088, ferdinandsu, 1183710221\}@stu.hit.edu.cn, \{tzy\_hit, xhc, rainy\}@hit.edu.cn}}
\maketitle              
\begin{abstract}

The dramatic development of cloud and edge computing allows for better Quality of Service (QoS) in many scenarios by deploying services on cloud and edge servers. Microservice technology is also adopted in these scenarios to decompose complex business logic into many small independent services. Meanwhile, as microservice systems continue to grow, providing stable QoS in these systems becomes a challenge, and many different approaches have been proposed for stable QoS. However, the microservice systems used in the experiments of these work have problems such as the low number of services and a single type of service. Therefore, we developed the open-source benchmark microservice system \SSName{} with 20+ services, including database, front-end, business logic, data processing, and artificial intelligence services in the \ScenarioName{}. Measuring tools are provided to measure the service properties to help researchers prepare experimental data, and the service properties pre-measured are also presented. Meanwhile, the fulfillment of benchmark requirements is detailed, and the results show that our \SSName{} meets the requirements of a benchmark system in research. Moreover, instructions are given to describe adopting our system in service computing as examples.

\end{abstract}

\section{Introduction}

Cloud computing and edge computing are adopted into many scenarios~\cite{chen2020iot} because of the adequate resources in the cloud and low delay in the edge, like the Internet of Things and smart city, and the size of the system also increases. Meanwhile, microservice technology is introduced into these scenarios with the increasing complexity of business logic. As a result, traditional applications are decomposed into several independently developed and deployed microservices~\cite{Microservices2018}, making the service system more complex, and it becomes challenging to keep stable Quality of Service (QoS) when the user requirements change. Therefore, plenty of research has been conducted on the service placement~\cite{hu2019optimizing}, scheduling~\cite{8651324}, etc. to improve QoS in mobile edge and cloud computing.

Meanwhile, problems with the experimental service sets raise when conducting research in these areas. Currently, there are two main types of experiments in these areas: the experiments with the properties of the services, including the service name, resource usage, and capability, and the experiments with runnable services. The former includes algorithm experiments and simulation experiments. They use the service's properties as input and do not need to run the service; the latter includes physical experiments. The service needs to be deployed to the server to observe the actual running effect besides the service's properties as input. To ensure the validity of the experiments in these work, the selection of the service set in the experiments is crucial. However, there is no high-quality service set with rich functionality and comprehensive scene coverage.

In the work~\cite{yu2019joint,9740415}, the experiments with the properties of the services were conducted, and the service sets were generated by themselves instead of measuring the service properties at runtime. The self-generated service properties make it difficult to compare the algorithms' performance fairly because of the different service properties used in different work. Furthermore, the generated data cannot represent the service properties at runtime, resulting in conclusions that mismatch the performance in the real world.

In other work~\cite{liu2021empirical}, the open-sourced service sets like Train-Ticket~\footnote{https://github.com/FudanSELab/Train-Ticket} were used, and experiments were conducted in a simulation environment or real servers. However, the used service sets focus on a single application scenario, making them offer few categories of functionality. For example, the Sock Shop~\footnote{https://github.com/microservices-demo/microservices-demo} only provides the functions about selling socks, which is not enough to cover the diversity of user requirements. The single application service set also makes it challenging to combine many service sets into one due to the non-interoperable business logic in different applications. Meanwhile, the service category in existing service sets cannot cover the application scenarios, like artificial intelligence (AI) and streaming data processing services, which are common in IoT scenarios.

A high-quality service set with rich functionality and comprehensive scene coverage is needed to solve these problems related to the experimental service sets. In this work, we designed and implemented a benchmark microservice system. It describes a simplified \ScenarioName{}, including sensor simulation services, AI-based algorithm services, streaming data processing services, front-end services, and other business logic services. The properties of services are also measured by experiments in a real container-based environment and provided with the measuring tools. To ensure the reliability of \SSName{} as a benchmark system, the benchmark system requirements in~\cite{aderaldo2017benchmark} are adopted to analyze \SSName{}, and instructions are also provided to show how to adopt the service set in the service placement and offloading problems.

The rest of the paper is organized as follows. Section~\ref{sec:background} shows the motivation. Section~\ref{sec:service_set} presents the details of the service set. Section~\ref{sec:application} analyzes the benchmark system requirements and introduces the instructions for adopting the service set in some research problems. Section~\ref{sec:conclusion} concludes the paper and outlines possible future work.

\section{Motivation}~\label{sec:background}

There exist various research problems in the field of microservice, and different microservice systems are adopted in experiments. By analyzing existing works and the service set used in their experiments, we present the characteristics that a high-quality benchmark microservice system should have for experiments.

\subsection{Microservice Research Problems}

We summarize some research problems in the field of microservice, and the collections of microservice systems used in these experiments are shown in Table~\ref{table:table-serviceQuestion}. There are three types of microservice systems used in existing works: enterprise, simulation, and runnable microservice system. The enterprise microservice system stands for the non-open sourced enterprise microservice systems used in research. The simulation microservice system means the generated microservice systems with key properties like resource usage and capabilities, and there is no implementation. Meanwhile, the runnable microservice system shows the runnable and open-sourced microservice systems developed by the researchers as benchmark microservices.

\begin{table}[ht]
\centering
\caption{Microservices Research}
\begin{tabular}{|c|ccc|}
\hline
\multirow{2}{*}{Research objectives} & \multicolumn{3}{c|}{Microservice System}             \\ \cline{2-4} & \multicolumn{1}{c|}{Enterprise} &  \multicolumn{1}{c|}{Simulation} &  Runnable \\ \hline

Design/Migration & \multicolumn{1}{c|}{\cite{gouigoux2017monolith}}   &  \multicolumn{1}{c|}{ -----}   &   \cite{al2020metrics,ren2018migrating,hippchen2017designing}               \\ \hline

Component/Framework & \multicolumn{1}{c|}{\cite{schmitt2020arcade}}   &  \multicolumn{1}{c|}{ -----}   &   \cite{he2020optimal,wang2021epf4m,liu2020mv4ms,xu2019microservice}               \\ \hline

Deployment/Placement/Evolution & \multicolumn{1}{c|}{\cite{ray2020proactive}}   &  \multicolumn{1}{c|}{ \cite{9740415,yu2019joint,hu2019optimizing,he2019re,chen2020iot,leitner2016modelling}}   &   \cite{he2020optimal,sampaio2017supporting,hexiang2021ruanjianxuebao}               \\ \hline


System Delay & \multicolumn{1}{c|}{\cite{de2019architectural}}   &  \multicolumn{1}{c|}{ \cite{de2020towards,ntentos2020assessing}}   &   \cite{liu2021empirical,pigazzini2020towards,ntentos2021evaluating}               \\ \hline

Scheduling/Offloading & 
\multicolumn{1}{c|}{\cite{lin2019ant}} & 
\multicolumn{1}{c|}{\cite{wang2020elastic,zhao2020distributed,fan2021multi,samanta2019battle,gedeon2019microservice}}  & 
\multicolumn{1}{c|}{\cite{bao2019performance}}  \\ \hline

\end{tabular}
\label{table:table-serviceQuestion}
\end{table}

For \textbf{Design/Migration}, it focuses on the design of microservice systems and the migration from the traditional monolithic architecture to microservice architecture by considering various indicators and standards. In the work of \cite{al2020metrics}, researchers use indicators like service granularity and cohesion to design microservices. Therefore, this research needs to conduct experiments with runnable microservice systems to detect whether their approaches are reasonable.

For \textbf{Component/Framework}, researchers enhance the functionality of microservices like traffic routing and registry by developing frameworks and components. For example, as in \cite{he2020optimal}, the multi-version dependency model of microservices is described through a Programming Framework, and service requests are routed through a Gateway Component.
These works require runnable microservice systems for testing and experimentation.

For \textbf{Deployment/Placement/Evolution}, it generates an optimized microservice deployment/re-deployment scheme to improve the QoS and reduce system costs, like average response time, deployment cost, and resource utilization in complex distributed scenarios (cloud, edge, fog, local clusters, etc.). To improve the persuasiveness of experimental results, the experiments need to consider the complex runtime environment of the microservice system, including the microservices with complex dependencies. As shown in Table~\ref{table:table-serviceQuestion}, in addition to choosing runnable experimental environments and systems, most of the research also chooses the simulation microservice systems.


For \textbf{System Decay}, the long-term evolution and iteration of software cause the problem of `Architecture Decay'\cite{2013Controlling}, and its specific manifestation is called bad smell. It is generally based on the microservice system of the enterprise, or the injection or elimination of bad smell in the runnable microservice systems, while some more abstract bad smells (such as service coupling) adopt the simulation microservice systems.


For \textbf{Scheduling/Offloading}, it schedules/offloads tasks to the best server by solving optimization problems, which helps to reduce cluster latency\cite{fan2021multi} and improve service reliability and availability\cite{zhao2020distributed}. This type of research seldom pays attention to the business logic and code implementation of microservices, and complex cloud edge environments are not easy to build, and most researchers use simulated microservice systems for experiments.

In summary, microservices have achieved great success in the industry, but the microservices of enterprises are usually not open sourced. In addition, the microservice system in the enterprise's production environment needs to ensure stability, so many academic researchers cannot directly use enterprise-level services. Therefore, a benchmark microservice system is a choice for most research.

\subsection{Open-Source Microservice Set}

We searched the microservice sets used in each microservice research and checked whether it is open-sourced on GitHub, and we excluded service sets with less than 7 microservices. When calculating the number of microservices, we also excluded Database Service, Message Middleware Service, Infrastructure Service (e.g., Service Registry, Gateway), and the rest that have nothing to do with business functions service. The results are summarized in Table~\ref{table:table-opnesource}. The table shows the number of microservices, service categories (e.g., compute-intensive services, IO services, database-connected services), interaction modes (e.g., synchronous, asynchronous, message queue), and the main language.

According to the results, most of the existing open-source microservice sets conform to the basic microservice design principles. However, there exist the following problems: 

\begin{itemize}
	\item[$\bullet$] \textbf{Small number of services}: Some of the service sets in the table are limited to about 10 services. The size of the solution space is related to the size of the microservice system, which affects the algorithms' performance. Small service sets cannot adequately verify the performance of the algorithms. Moreover, small service sets cannot represent scenarios with growing scales.

	\item[$\bullet$] \textbf{Simple dependencies between services}: The dependency relationship between services is relatively simple. The reusability emphasized in the microservice design principles leads to complex dependencies, which means one microservice can be reused in multiple applications.
	
	\item[$\bullet$] \textbf{Fewer types of service functions}: Many open source systems have a single business and similar business logic (e-commerce systems are the most common microservice sets. Most of them are connected to the database and used for business processing (Train Ticket), which does not meet the needs of complex scenarios such as edge intelligent AI, data processing, and sensor data collection. Because different service functions have different characteristics, the experiment service sets should have various function types to represent the real scenarios.
\end{itemize}


\begin{table}[ht]
\centering
\caption{Open-source Microservice set in researches}
\begin{tabular}{|c|c|c|c|c|}
\hline
\multicolumn{1}{|c|}{System Name} & \# Servive &  Category  & Interaction Mode & Main Languages\\ \hline
Train Ticket  & 40 &DATABASE  &Sync, Async & Java,Python,JavaScript \\ \hline
Sock Shop & 8 &  DATABASE  & Sync,Queue&Python,HCL           \\ \hline
DeathStarBench\tablefootnote{https://github.com/delimitrou/DeathStarBench}&         8    & DATABASE  & Sync  &   C, Lua       \\ \hline
Sitewhere\tablefootnote{https://github.com/sitewhere/sitewhere}   &      19    & DATABASE  & Sync,Queue   &  Java \\ \hline
micro-company\tablefootnote{https://github.com/idugalic/micro-company}&    7   &          DATABASE        &      Sync,Queue         & JavaScript, Java \\ \hline
Staffjoy\tablefootnote{https://github.com/Staffjoy/v2} & 10 & DATABASE & Sync,Queue & JavaScript,Go \\ \hline
eShop\tablefootnote{https://github.com/dotnet-architecture/eShopOnContainers} & 14 & DATABASE & Sync,Queue & C\#,JavaScript \\ \hline
\end{tabular}
\label{table:table-opnesource}
\end{table}

\subsection{Ideal Microservice Set}

Therefore, as a microservice system that can satisfy the academic community, it should have the following characteristics.

\begin{itemize}
    \item[$\bullet$] The microservice system must conform to the design principles of microservices. Each microservice is responsible for a certain business or function and has an independent repository and version iteration process.
    \item[$\bullet$] The number of services must be sufficient and the dependencies between them must be complex enough.
    \item[$\bullet$] The service categories in the system should be varied, including database connection services, aggregation services, computing-intensive services, and AI computing services.
\end{itemize}

To satisfy the above conditions, we developed a microservice system named \SSName{} detailed in Section~\ref{sec:service_set}.

\section{Service Set}~\label{sec:service_set}

This section describes the design principles of \SSName{}\footnote{https://anonymous.4open.science/r/RescueServiceOverview-ReviewOnly}. It also gives an overview of the microservice system and details key properties of each service. Meanwhile, a benchmark tool that aims to measure the runtime metrics for each service is presented, and experiment data pre-measured is provided.

\subsection{Overview}


The design of \SSName{} adheres the following principles:

\begin{itemize}


    \item[$\bullet$] \textbf{User-defined routing}: \SSName{} provides a collection of services without components such as gateways so that researchers can control routing rules in issues like service composition, orchestration, and evolution. Users can freely use eureka, k8s, and istio to combine and autonomously perform fine-grained service traffic scheduling.
    
    \item[$\bullet$] \textbf{Synchronous requests}: All the requests in \SSName{} are synchronous. The experiment results cannot be affected because asynchronous and synchronous requests can be described in the same models~\cite{8651324}, and the microservice's performance keeps the same for asynchronous and synchronous requests~\cite{jindal2019performance}.
    
    \item[$\bullet$] \textbf{Representative technologies}: \SSName{} is implemented using some representative and popular methods, such as MySQL, RabbitMQ, Spring, Docker, Kubernetes, and etc.
    
    \item[$\bullet$] \textbf{Various service categories}: \SSName{} contains various service categories, like database services, aggregation services, computing-intensive services, AI computing services, and streaming data processing services.
\end{itemize}

\begin{sidewaysfigure}[htbp]
    \includegraphics[width=\textwidth]{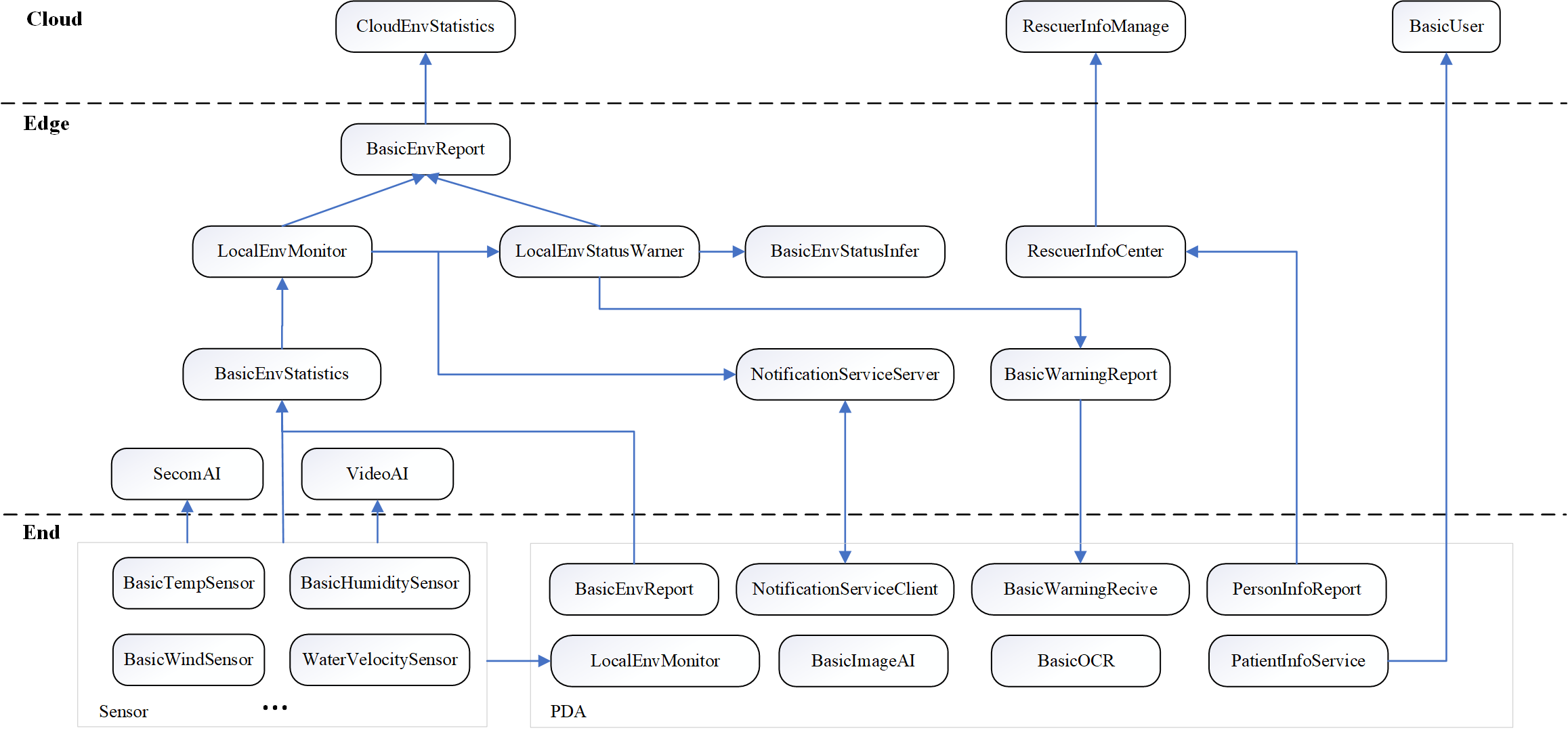}
    \caption{Service topology}\label{fig:topology}
\end{sidewaysfigure}

Based on the above principles, this paper designs and implements \SSName{}, which consists of more than 20 microservices, as shown in Figure~\ref{fig:topology}. \SSName{} is designed based on \ScenarioName{}, including environmental status collection, disaster warning, and rescuer management. These microservices run in different environments, including the cloud, the edge, and the endpoint sensors. It should be noted that the algorithms are not our primary concern, and the algorithms' performance cannot be ensured.

\subsection{Service Design}

This subsection describes the key properties of each microservice in \SSName{}. As shown in Table~\ref{table:serviceset}, microservices are classified into different types based on their functionality, such as AI and Database. in addition, disk usage and abilities are described. Because the service resource usage depends on the hardware and runtime environment, it is not included in the table.


\begin{table}[h]
\centering
\caption{Description for Services in \SSName{}}
\label{table:serviceset}
\begin{tabularx}{\linewidth}{|l|l|l|X|} 
\hline
Service & Category & DISK  & Description \\ \hline
BasicUser  & DATABASE &  150MB &Provide basic information of all users   \\ \hline
NotificationServer  & DATABASE &  151MB & Provide communication, broadcasting, and other functions between users
\\ \hline
NotificationClient &  Function &  131MB & Client side of the notification service  \\ \hline
MedicalGuidance  & DATABASE &  151MB & Allow upload/download video for medical guidance \\ \hline
PatientInfoService &  Function &  133MB & Provide patient information  \\ \hline
BasicMaterialManage & DATABASE & 117.99MB & Rescuer material management. \\ \hline
RescuerInfoManage & DATABASE & 110.42MB & Rescuer info management. \\ \hline
RescuerInfoCenter & DATABASE & 110.98MB & Rescuer/Update the status of rescuers. \\ \hline
LocalEnvMonitor & FUNCTION & 110.98MB & Collect data of sensors and reporting. \\ \hline
BasicEnvReport & FUNCTION & 130MB & Report environmental status and data \\ \hline
BasicEnvStatusInfer & AI & 131MB & Infer state through environmental data \\ \hline
BasicWarningReport & FUNCTION & 135MB & Broadcast warning \\ \hline
BasicEnvStatistics & FUNCTION & 110.04MB & Collect sensor data to edge \\ \hline
CloudEnvStatistics & DATABASE & 151MB & Collect data from the edge \\ \hline
LocalEnvStatusWarner & DATABASE & 151MB & Receive data information and analyze status \\ \hline
BasicTempSensor & SENSOR & 91.01MB & Virtual temperature sensor \\ \hline
BasicHumiditySensor & SENSOR & 91.01MB & Virtual humidity sensor \\ \hline
BasicWindSensor & SENSOR & 91.22MB & Virtual wind sensor \\ \hline
WaterVelocitySensor & SENSOR & 91.22MB & Virtual water velocity sensor \\ \hline
BasicImageAI & AI & 1.57GB & Image Recognition \\ \hline
BasicOCR & AI & 256.70MB & Optical character recognition \\ \hline
SecomAI & AI & 458.39MB & Predict production result with monitored data \\ \hline
VideoAI & AI & 401.07MB & Object motion detection \\ \hline
\end{tabularx}
\end{table}


Several experiments were conducted to measure the runtime performance metrics of individual microservices, including service capabilities and resource consumption under different loads. For this purpose, a benchmark tool\footnote{https://anonymous.4open.science/r/RescueServiceOverview-ReviewOnly} is developed that automatically deploys the service, simulates user requests, collects request results, and monitors the resource consumption of the service during the request process. During the experiments, individual microservices were sandboxed to facilitate the measurement of each service's performance metrics individually, with the following results.

All the experiments were conducted on a CentOS 7 Server with CPU Intel Xeon Gold 5120 and 160GB RAM, and Kubernetes V1.19.2 was used to deploy the services. Based on the CPU limitation feature of Kubernetes, services were tested with CPU limitations ranging from 200m and 1000m, and the number of requests processed per second was recorded to describe the service capabilities. Experiments can also be conducted w.r.t. RAM usage with the measuring tools. However, we noticed that different ram limitations lead to the same service capabilities for the services in \SSName{}. Thus, we only measured the service capabilities w.r.t. the CPU limitation.

It should be noted that the service capabilities can be different on different machines, and all the service capabilities should be measured in the target machines with the measuring tools for accuracy. The experiment results are uploaded to the repository with the service set \SSName{}\footnote{https://anonymous.4open.science/r/RescueServiceOverview-ReviewOnly}.

\begin{figure}
   \centering
   \subfigure[Service BasicUser]{
    \includegraphics[width=0.46\textwidth]{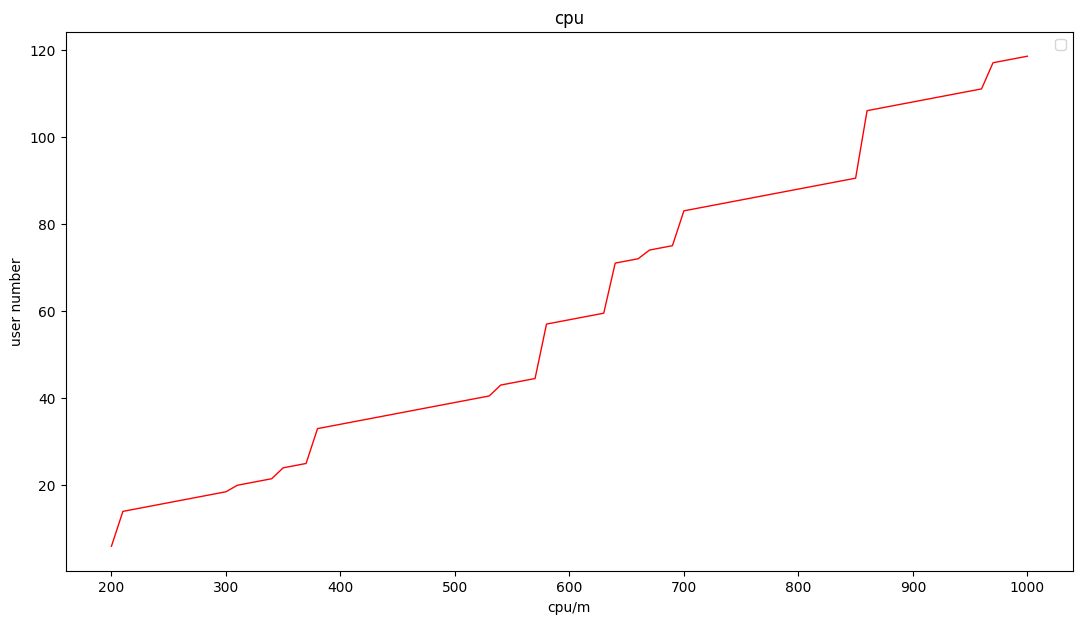}\label{subfig:basic_user}
   }
   \subfigure[Service NotificationServer]{
    \includegraphics[width=0.46\textwidth]{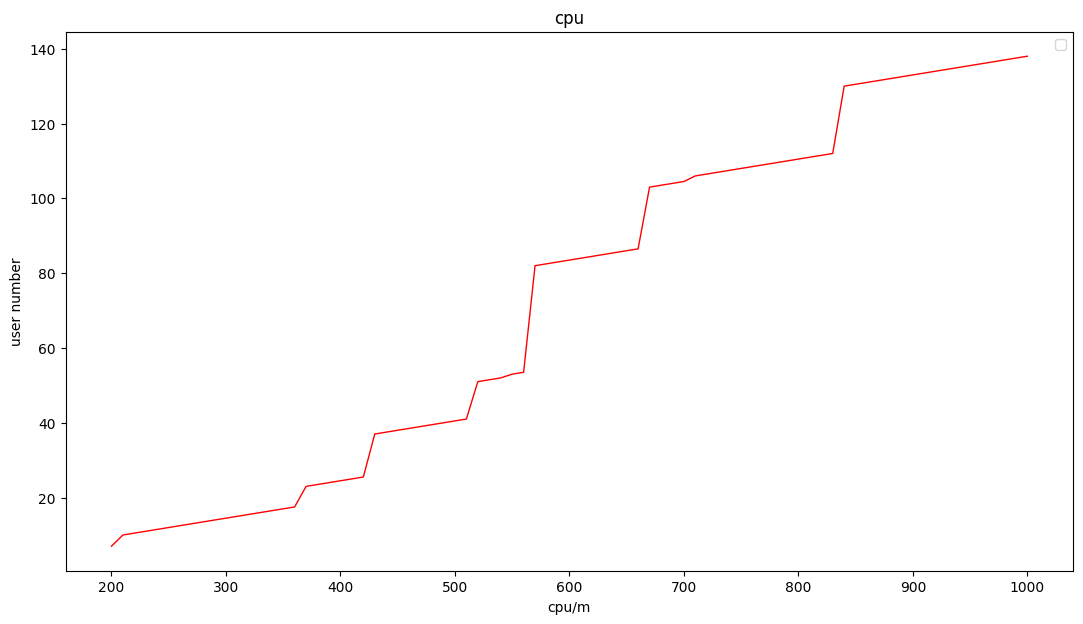}\label{subfig:notificationServer}
   }
   \caption{Service capabilities w.r.t. cpu limit for BasicUser and NotificationServer}\label{fig:experiments}
\end{figure}

\section{Application Scenarios}~\label{sec:application}



In this section, to ensure \SSName{} can fit the requirements of the benchmark experiments in various research problems, the benchmark microservice system selection requirements proposed in ~\cite{aderaldo2017benchmark} are employed, and the benchmark requirements are analyzed for our \SSName{}. Meanwhile, to demonstrate how to use the \SSName{} in various microservice-related research problems, we pick the service placement and offloading problems as examples. Finally, with brief problem definitions, the instructions for using \SSName{} are given.

\subsection{Benchmark requirements for Microservice System}

The work~\cite{aderaldo2017benchmark} proposed a set of requirements that is useful in selecting a benchmark microservice system. The benchmark requirements are divided into three categories: architecture requirements, DevOps requirements, and general requirements, and there are 12 requirements in total, as shown in Table~\ref{table:requirement}.

We evaluated the \SSName{} using the criteria in Table~\ref{table:requirement}, and the results are shown in Table~\ref{table:assessResult}. The \SSName{} satisfies R1--R9 of the 12 benchmark requirements. 
As the first version of the \SSName{} system, limited technologies are used, and it currently lacks version iterations, which can be achieved through continuous iterations. Thus, its performance in General is poor, and there are problems such as single technology and insufficient community influence.

\begin{table}[htp]
\centering
\caption{Benchmark Requirements for Microservice System.}
\label{table:requirement}
\begin{tabularx}{\textwidth}{|l|p{3cm}|X|} 
\hline
Context                       & Requirement                                       & Assessment Rationale                                                                                                                 \\ 
\hline
\multirow{2}{*}{Architecture} & R1: Explicit Topological View                     & The benchmark should provide an explicit description of its main service elements and their possible runtime topologies.~            \\ 
\cline{2-3}
                              & R2: Pattern-based architecture                    & The benchmark should be designed based on well-known microservices architectural patterns.                                           \\ 
\hline
\multirow{7}{*}{DevOps}       & R3: Easy access from a Version Control Repository & The benchmark's software repository should be easily accessible from a public version control system.~                               \\ 
\cline{2-3}
                              & R4: Support for Continuous Integration            & The benchmark should provide support for at least one continuous integration tool.                                                   \\ 
\cline{2-3}
                              & R5: Support for Automated Testing                 & The benchmark should~provide support for~at least one automated test tool.                                                           \\ 
\cline{2-3}
                              & R6: Support for Dependency Management             & The benchmark should~provide support for~at least one dependency management tool.                                                    \\ 
\cline{2-3}
                              & R7:~Support for Reusable Container Images         & The benchmark should~provide reusable container images for at least one container technology.                                        \\ 
\cline{2-3}
                              & R8:~Support for Automated Deployment              & The benchmark should~provide support for~at least one automated deployment tool.                                                     \\ 
\cline{2-3}
                              & R9:~Support for Container Orchestration           & The benchmark should~provide support for~at least one container orchestration tool.                                                  \\ 
\hline
\multirow{3}{*}{General}      & R10: Independence of Automation Technology        & The benchmark should~provide support for multiple technological alternatives at each automation level of the DevOps pipeline.        \\ 
\cline{2-3}
                              & R11: Alternate Versions                           & The benchmark should~provide support for alternate implementations in terms of programming languages and/or architectural decisions.  \\ 
\cline{2-3}
                              & R12: Community Usage  Interest                    & The benchmark should be easy to use and attract the interest of its target research community.                                       \\
\hline
\end{tabularx}
\end{table}

\begin{table}
\centering
\caption{Assess Results of \SSName{} System.}
\label{table:assessResult}
\begin{tabularx}{\textwidth}{|l|l|X|} 
\hline
Requirement & Satisfy & Assessment Results                                                                         \\ 
\hline
R1          & Yes                  & Overall services topology clearly described                                                \\ 
\hline
R2          & Yes                  & Service Discovery, Database per Service, Messaging                                          \\ 
\hline
R3          & Yes                  & Source code publicly available at GitHub                                                   \\ 
\hline
R4          & Yes                  & Uses GitLab CI for CI                                                                      \\ 
\hline
R5          & Yes                  & Use GitLab for testing                                                                     \\ 
\hline
R6          & Yes                  & Use Maven and NPM for Deployment Manage                                                                   \\ 
\hline
R7          & Yes                  & Reusable Docker images                                                                     \\ 
\hline
R8          & Yes                  & Uses Kubernetes for deployment                                                             \\ 
\hline
R9          & Yes                  & Uses Kubernetes for orchestration                                                          \\ 
\hline
R10         & No                  & No support for alternative technologies                                                   \\ 
\hline
R11         & No                  & No alternative versions                                                                    \\ 
\hline
R12         & No                  & No previous usage found but recent GitHub activity indicates potential community interest  \\
\hline
\end{tabularx}
\end{table}

Although the \SSName{} cannot satisfy R9, R10 and R11, it fully supports the Architecture and DevOps requirements. 
Many researches in the field of edge computing and fog computing require the service system to be automatically deployed according to the generated deployment scheme, and to monitor the performance indicators of each service for verification. The \SSName{} system satisfies these requirements well.
Moreover, compared with other open-source microservice systems, \SSName{} system contains more types of technologies which are applied in industry and more real-life application scenarios.
For example, the system includes scenarios such as sensor data reporting, environmental condition prediction, and data streaming transmission. Compared with other open-source microservice systems, there are explicit cloud-edge-end application scenarios in the \SSName{}, which are more suitable for use as experimental data sets in the fields of edge computing and fog computing.

Above all, \SSName{} microservice system can support researchers to get reliable experimental results because of various functional categories. Furthermore, the open-source service set ensures the reproducibility of experiments so that researchers can compare the performance of different approaches in a runnable service system.

\subsection{Potential Scenario: Service Placement}


The performance and resource utilization of a microservice system depend on the deployment location and actual physical environment of the microservices that make up it.
Therefore, how to find a service placement solution that optimizes the performance of the microservice system is a classic research problem in the cloud/edge computing.

Service placement problem consists of three parts: infrastructure model, application model, and constraint. The infrastructure model is an abstraction of the actual physical environment which consists of a set of devices (end devices, such as sensors) and a set of resources with computational or storage capabilities (edge nodes and cloud data/compute centers). 
The application model describes the service system, which can be a monolithic or microservice system. 
The constraints include service constraints, resource constraints, QoS constraints, and more.

In summary, the service placement problem is defined as finding a mapping pattern to map the application model to the infrastructure model. Meanwhile, the mapping pattern needs to ensure the system's availability and improve the system's QoS as much as possible on the condition of meeting the constraints. The benchmark microservice system \SSName{} could be useful to conduct the experiments in real world.



\subsection{Potential Scenario: Offloading Decision}


With the development of IoT, edge computing, and fog computing, offloading has received more and more attention. In fog computing, there are a large number of end computing devices that are interconnected. Due to the limitation of resources and computing ability, these end computing devices need to send requests to the cloud for processing, leading to high latency and the inability to meet the low latency requirements of some real-time applications. In order to solve this problem, the idea of offloading was born. On the one hand, limited resources are solved by shifting the computing workload to other devices with better resources. On the other hand, the offloading decision is made based on the QoS parameters of the service to select a suitable location for the computing device to shift the computing workload to reduce the request delay. In summary, the definition of the offloading decision problem is to select the appropriate location of the computing device to shift the computing workload, reducing the delay in meeting the requirements of real-time tasks.

\subsection{Instructions for Use}
Service placement and offloading decision problems have similar requirements for experimental data sets. The data sets are required to provide resource information for each service, including edge application scenarios, support automated deployment services and custom routing, etc. The difference between the two mainly in deployment schemes, routing, and monitor metrics. The \SSName{} system can meet these requirements, such as providing the resource requirements of each service, including sensor transmission data and early warning services, and supporting custom routing through eureka, etc.

The following steps introduce how to use the \SSName{} system to conduct physical experiments for research like service placement and offloading. 

\begin{enumerate}
    \item [Step 1] (Configure the runtime environment) Install the service orchestration tool Kubernetes as the runtime environment of the \SSName{} microservice system.
    \item [Step 2] (Deploy the registry center) Deploy a registry center (such as Eureka) for the \SSName{} system ensuring that each service can be registered and discovered normally.
    \item [Step 3] (Deploy and run services) For each service in the \SSName{} system, configure its image repository and resource. Take use of tools and scripts provided by the \SSName{} system, deploy the image of each service to the specified node by means of Kubernetes Pod through deployment scheme, and ensure the service is up and running.
    \item [Step 4] (Monitor performance metrics) After the system is running, monitor the QoS metrics of the system with the help of some open source monitoring tools.
\end{enumerate}

\section{Conclusion and Future Work}~\label{sec:conclusion}

This paper presented an open-source microservice set \SSName{} as the benchmark microservice system for experiments in edge and cloud computing. By analyzing the microservice sets used in many research, we addressed the characteristics that a high-quality benchmark microservice set should have for experiments, including the microservice set size, complex service dependencies, and types of functions. The benchmark requirements for \SSName{} are also analyzed, and the result shows our \SSName{} can support researchers to conduct reproducible experimental studies. Instructions are also presented to guide users how to use \SSName{} in different research areas.

The possible future work includes extending the proposed \SSName{} with more microservices. In addition, version control should also be addressed. Some works that focus on microservice system evolution can benefit from the version history of each microservice.




%
%
%
\bibliographystyle{splncs04}
\bibliography{ref.bib}

\end{document}